\documentclass{article}

\usepackage{multirow} 
\usepackage{cite} 
\usepackage{wrapfig}

    \usepackage[preprint]{neurips_2022}

\usepackage[utf8]{inputenc} 
\usepackage[T1]{fontenc}    
\usepackage{hyperref}       
\usepackage{url}            
\usepackage{booktabs}       
\usepackage{amsfonts}       
\usepackage{nicefrac}       
\usepackage{microtype}      
\usepackage{xcolor}         
\usepackage{graphicx}
\usepackage{subfigure}

\title{Preprocessing Enhanced Image Compression for Machine Vision}

\author{%
  Guo Lu \\
  Beijing Institute of Technology, China \\
  \texttt{sdluguo@gmail.com} \\
   \And
   Xingtong Ge  \\
   Beijing Institute of Technology, China \\
   \texttt{xingtong.ge@gmail.com} \\
   \AND
   Tianxiong Zhong \\
   Beijing Institute of Technology, China \\
   \texttt{inkosizhong@gmail.com} \\
   \And
   Jing Geng \\
   Beijing Institute of Technology, China \\
   \texttt{janegeng@bit.edu.cn} \\
   \And
   Qiang Hu \\
   ShanghaiTech University, China \\
   \texttt{huqiang@shanghaitech.edu.cn} \\
}

\begin{document}

\maketitle

\begin{abstract}
Recently, more and more images are compressed and sent to the back-end devices for the machine analysis tasks~(\textit{e.g.,} object detection) instead of being purely watched by humans. 
However, most traditional or learned image codecs are designed to minimize the distortion of the human visual system without considering the increased demand from machine vision systems. In this work, we propose a preprocessing enhanced image compression method for machine vision tasks to address this challenge. Instead of relying on the learned image codecs for end-to-end optimization, our framework is built upon the traditional non-differential codecs, which means it is standard compatible and can be easily deployed in practical applications. Specifically, we propose a neural preprocessing module before the encoder to maintain the useful semantic information for the downstream tasks and suppress the irrelevant information for bitrate saving. Furthermore, our neural preprocessing module is quantization adaptive and can be used in different compression ratios. 
More importantly, to jointly optimize the preprocessing module with the downstream machine vision tasks, we introduce the proxy network for the traditional non-differential codecs in the back-propagation stage. We provide extensive experiments by evaluating our compression method for two representative downstream tasks with different backbone networks. Experimental results show our method achieves a better trade-off between the coding bitrate and the performance of the downstream machine vision tasks by saving about 20\% bitrate.  
\end{abstract}

\section{Introduction}

Due to the successful applications of deep neural networks, the machine vision tasks such as detection and classification have made a lot of progress in recent years~\cite{tian2019fcos,ren2015faster,lin2017RetinaNet,redmon2018yolov3,bochkovskiy2020yolov4,he2016resnet, zhang2020resnest, xie2017resnext}. Therefore, more and more images are captured at the front-end devices~(\textit{e.g.,} cameras) and sent to the back-end~(\textit{e.g.,} cloud servers) for machine analysis. According to the report from Cisco~\cite{Cisco}, the percentage of the connections from this machine-to-machine scenario~(\textit{e.g.,} video surveillance) will be up to 50\% in the future. 
Therefore, how to reduce the transmission bitrate while maintaining the performance for the downstream machine vision tasks has become a challenge for the image compression field.

\begin{figure}
  \centering
  \includegraphics[width=1\columnwidth]{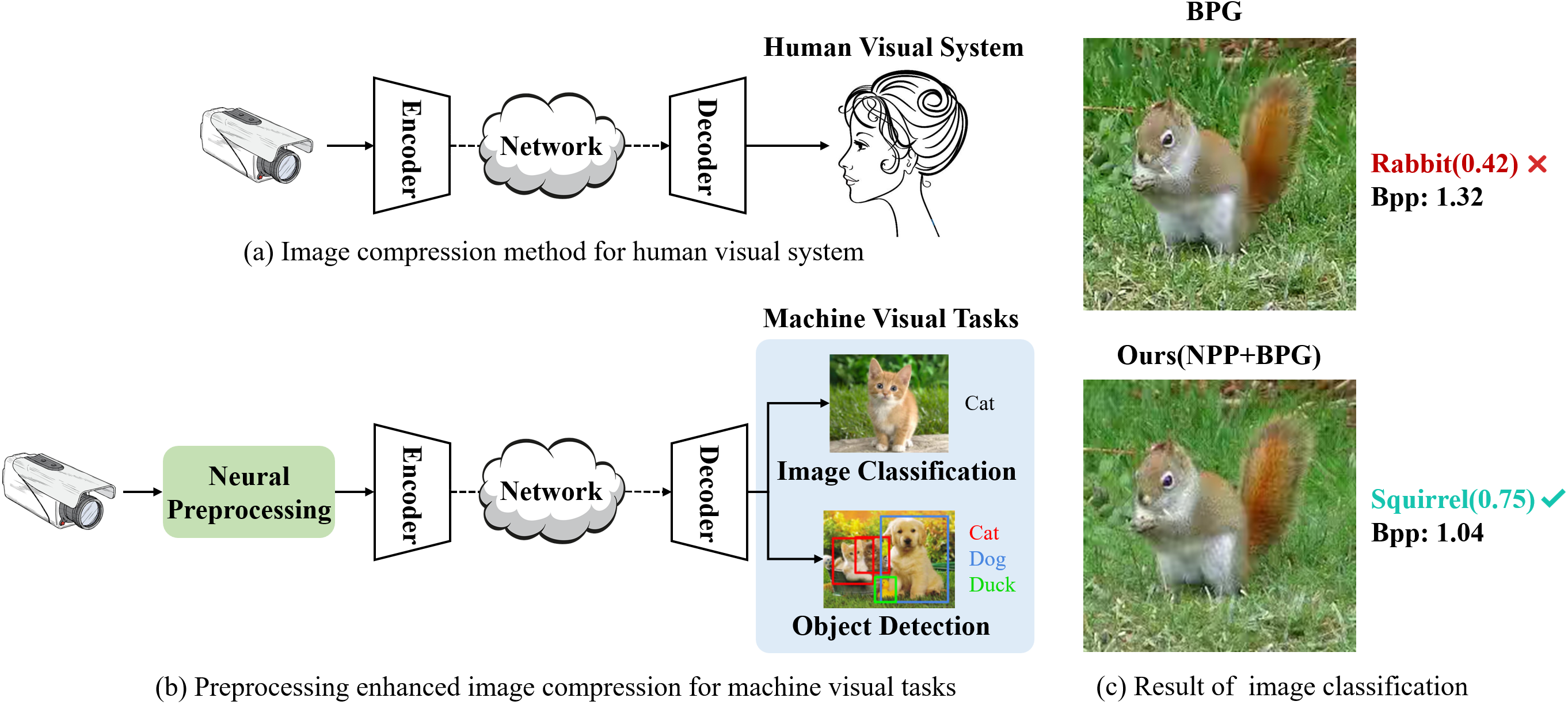} 
  \caption{(a) Image compression method for human visual system. (b) Our proposed preprocessing enhanced image compression for machine vision tasks. (c) Image classification results for the image from the BPG codec and ours(NPP+BPG).}
  \label{fig:conception}
\end{figure}

Unfortunately, although several traditional image compression standards, such as JPEG~\cite{wallace1992jpeg} and BPG~\cite{BPG}, have been proposed in the past decades, they are designed to minimize the compression distortion for the human visual system~(\textit{e.g.}, PSNR) instead of the machine vision tasks~(see Fig.~\ref{fig:conception}(a)).
More importantly, most compression standards are non-differential, which cannot be jointly optimized with the neural network based machine analysis methods. Therefore, the existing compression-then-analysis pipeline with the traditional codecs may not be optimal when we mainly focus on the performance of the downstream machine analysis.

Recently, learned image compression methods~\cite{balle2016end,balle2018variational,minnen2018joint} start to gain a lot of attention. 
Several approaches~\cite{torfason2018towards,akbari2019dsslic,hu2020towards} also have tried to jointly optimize the learned compression methods with the downstream analysis tasks. 
However, the computational complexity for the learned image compression is usually high, and the standardization is not finalized; therefore, the massive deployment of learned compression approaches is unlikely to happen soon, which means these approaches~\cite{torfason2018towards,akbari2019dsslic,hu2020towards} may not be feasible in practical applications. 

To address these challenges, we propose a preprocessing enhanced image compression framework for machine vision as shown in Fig.~\ref{fig:conception}(b). 
Our framework builds upon the traditional standard compatible image codecs and can be easily applied to the practical compression-then-analysis systems.
Specifically, we propose a neural preprocessing~(NPP) module before the traditional codec and the input image will be filtered before encoding. After that, the decoded image is used for the downstream vision tasks, like detection or classification.  
To enable the end-to-end optimization, we further introduce the proxy network for the traditional non-differential image codecs~(\textit{e.g.,} BPG) in the training stage, where the gradients of the proxy network are propagated to the neural preprocessing module. 
Therefore, the proposed preprocessing module will be optimized to maintain the meaningful semantic information and reduce the irrelevant information for machine vision tasks, which leads to a better trade-off between the coding bitrate and machine analysis performances~(see Fig.~\ref{fig:conception}(c)). 
Furthermore, the proposed neural preprocessing module is quantization adaptive and can be integrated into traditional codecs with different compression ratios. 
To demonstrate the superiority of our preprocessing enhanced image compression method, we perform extensive experiments on two representative machine vision tasks~(object detection and image classification) with different downstream backbone networks.
Experiments show that compared with the existing traditional codec like BPG~\cite{BPG}, the proposed approach can save about 20\% bitrate for object detection and image classification tasks with the same accuracy.

The main contributions of our work are summarised as follows,
\begin{itemize}
\setlength\itemsep{-.3em}
\item Building upon the traditional codec, we propose a neural preprocessing module to generate the filtered images, which the traditional codecs can effectively compress with high machine perception performance.
\item To enable an end-to-end optimization for a better trade-off between coding bitrate and machine perception performance, we introduce the learned proxy network to approximate the traditional codecs for the back-propagation in the training stage.
\item  Experimental results show our approach is general and the optimized NPP model for one specific scenario can be used for other codecs, downstream backbones, or even the other tasks.
\end{itemize}

\section{Related Works}
\subsection{Image Compression}
Many traditional image compression algorithms~\cite{wallace1992jpeg,skodras2001jpeg,BPG} have been proposed in the past decades. These methods are based on hand-craft techniques~(\textit{e.g.,} Discrete Cosine Transform) to reduce spatial redundancy. 
Recently, the learned image compression methods~\cite{balle2016end,johnston2017improved,toderici2017full,balle2018variational,minnen2018joint,zhu2021transformer,cheng2020resblock,chen2021nonlocal,xie2021invertible} have become popular. 
The mainstream methods~\cite{balle2016end,balle2018variational,minnen2018joint,zhu2021transformer} adopt an auto-encoder style network to convert the images to the latent representations, which are further encoded by entropy coding. 
For example, Ball\'e \emph{et al.}~\cite{balle2016end} proposed to use a convolutional neural network~(CNN) to learn non-linear transformation and  additionally introduce a hyper-prior network to model the probability distribution of the latent representations~\cite{balle2018variational}. 
Latest works~\cite{cheng2020resblock,chen2021nonlocal,xie2021invertible,zhu2021transformer} also propose to use more powerful transform networks, such as residual blocks~\cite{cheng2020resblock}, nonlocal layers~\cite{chen2021nonlocal}, invertible layers~\cite{xie2021invertible} and transformer~\cite{zhu2021transformer}. 
Although these approaches have achieved better compression performance, they are computationally expensive. 
More importantly, there is no coding standard for these learned compression methods, which cannot be massively deployed to practical applications. 

\subsection{Compression for Machine Vision}

Most existing image compression methods~\cite{wallace1992jpeg,BPG,balle2016end,balle2018variational} aim to reduce reconstruction distortion in terms of the human visual system and are optimized based on the pixel field metrics such as PSNR or MS-SSIM~\cite{wang2003msssim}.
With the development of deep learning, some studies~\cite{torfason2018towards,akbari2019dsslic,hu2020towards,li2021task,wang2021towards,yan2021sssic,fischer2021boosting} also focus on the joint optimization of the image compression and the downstream machine vision tasks. For example, Torfason \emph{et al.}~\cite{torfason2018towards} proposed to directly perform image understanding tasks, such as classification and segmentation, on the compressed representations produced by the learning-based image compression methods. 
Fischer \emph{et al.}~\cite{fischer2021boosting} introduced the feature loss to optimize the image compression network and achieved a better trade-off between bitrate cost and analysis accuracy. 
Wang \emph{et al.}~\cite{wang2021towards} proposed a scalable coding based compression for both face reconstruction and face analysis, where the base layer encodes the valuable features for analysis. In contrast, the enhancement layers encode the texture information for reconstruction.

However, most existing works have to rely on the learning based codecs to enable the end-to-end optimization, which may not be feasible in the practical application considering the mainstream codecs are traditional ones.
In contrast, our framework is built upon the traditional codecs and can also be end-to-end optimized through the proxy network. 

\subsection{Preprocessing}
In the past decades, several methods~\cite{xiang2016adaptive,vidal2017new,doutre2009color} have been proposed to use the preprocessing technique to improve the performance of the image and video compression algorithms. Most of these methods are based on the Just Noticeable Distortion~(JND) technique~\cite{yang2005just} and try to improve the perceptual quality of reconstructed frames.
For example, Xiang \emph{et al.}~\cite{xiang2016adaptive} proposed adaptive perceptual preprocessing by removing the information that is not perceptible to the human visual system. 
Vidala \emph{et al.}~\cite{vidal2017new} combined several adaptive filters to denoise the image for bitrate saving.

In recent years, several learning-based preprocessing methods have been proposed~\cite{Chadha_2021_CVPR,guleryuz2021sandwiched,talebi2021better,son2021enhanced}.
Chadha \emph{et al.}~\cite{Chadha_2021_CVPR} proposed a rate-aware perceptual preprocessing module for video coding. Onur \emph{et al.}~\cite{guleryuz2021sandwiched} proposed neural network based preprocessing and postprocessing modules to improve the compression performance of the traditional codecs. Talebi \emph{et al.}~\cite{talebi2021better} designed a pre-editing neural network on the JPEG method to improve the visual quality of reconstructed images.
In contrast, we propose using the neural network based preprocessing method to improve the compression performance in machine vision instead of the human visual system.

\begin{figure}
  \centering
  \includegraphics[width=1\columnwidth]{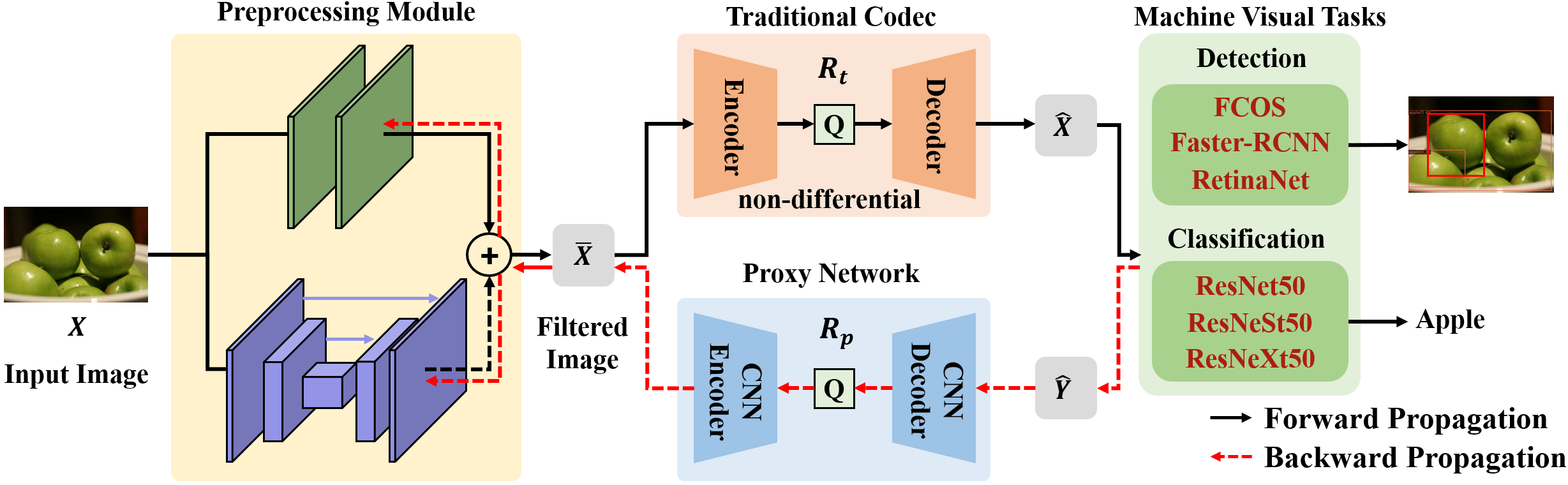}
  \caption{Overview of our preprocessing enhanced image compression for machine vision.}
  \label{fig:overview}
\end{figure}

\section{Proposed Method}
\subsection{Overview}
The overall architecture of our preprocessing enhanced image compression framework for machine vision is shown in Fig.~\ref{fig:overview}. 
The whole system aims to achieve a better trade-off between coding bitrate and the performance of the machine analysis task. 
Specifically,  we first feed the input image $X$ to the neural preprocessing module~(NPP) for non-linear transform and generate the filtered image $\bar{X}$, which is expected to maintain the critical semantic information. Then, $\bar{X}$ is encoded and reconstructed by a traditional codec, like BPG~\cite{BPG}. Finally, the decoded $\hat{X}$ is input to machine analysis networks, such as FCOS~\cite{tian2019fcos}. 

Since the traditional codecs maybe not be differential, the proposed preprocessing module cannot enjoy the benefits of the joint end-to-end optimization with the downstream machine analysis tasks. To solve this problem, we additionally introduce a learned image compression network as the proxy network for the traditional codec in the training stage and the gradients of the proxy network are propagated to preprocessing module~(see Section~\ref{sec:proxy} for more details). 
Here, we use BPG~\cite{BPG} as the traditional codec in our implementation.

Then the framework is optimized by using the following loss function,
\begin{equation}
    \mathcal{L}=R_t +  \lambda \mathcal{D}_{m} + \beta D_{pre}
\label{eq:loss1}
\end{equation}
where $\mathcal{D}_{m}$ and $R_t$ represents the loss of the downstream machine vision task based on reconstructed image $\hat{X}$ and coding bitrate from traditional codec, respectively.
$\lambda$ is a hyper-parameter used to control the trade-off.
In addition, to stabilize the training process, we also consider the distortion between the input image $X$ and the enhanced image $\bar{X}$, which is denoted as $D_{pre}$.  $\beta$ is the constant weight parameter.

\subsection{Neural Preprocessing}
\begin{figure}
  \centering
  \includegraphics[width=1\columnwidth]{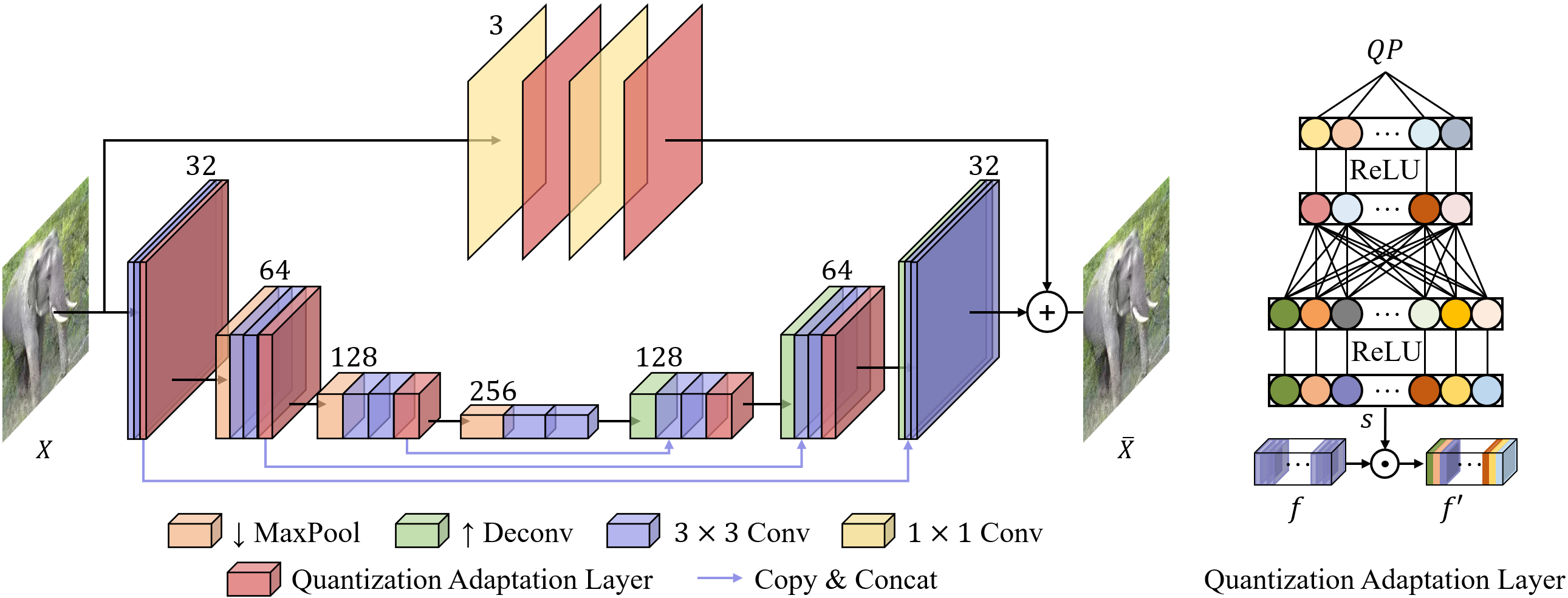}
  \caption{The implementation of our neural preprocessing module. The numbers represent the numbers of output channel for different operations. }
  \label{fig:preprocessor}
\end{figure}

As shown in Fig.~\ref{fig:preprocessor},  we provide the network architecture of our neural preprocessing module. 
Specifically, the original image $X$ is input into two parallel branches, where the first branch uses $1\times 1$ convolutional layers to learn non-linear pixel-level transform, and the second branch uses a U-Net~\cite{ronneberger2015unet} style network to extract the semantic information. 
The outputs of two branches are added together as the final filtered image $\bar{X}$, which preserves the useful texture and semantical information through both shallow and deep transforms.

Furthermore, considering the traditional codecs have different compression ratios~(\textit{i.e.,} quantization parameter), therefore, the neural preprocessing module is required to generate the optimal filtered image $\bar{X}$ for each compression ratio. 
Here, we propose a quantization adaptation layer for the neural preprocessing module, which leads to an adaptive preprocessing based on the quantization parameters in the codec.
As shown in Fig.~\ref{fig:preprocessor}, we integrate the quantization adaptive layer into the NPP module and scale the intermediate features for adaptive filtering. Specifically, based on the given quantization parameter~(QP) in the traditional codec, we use a 2-layer MLP network to generate the scale vector $s$ and the output feature $f'$ is the channel-wise multiplication product between input feature $f$ and generated scale vector $s$, \textit{i.e.,} $f' = f \odot s$. Based on this strategy, the intermediate features in the preprocessing module will be modulated based on the quantization parameter; therefore, our module will generate the optimal filtered image $\bar{X}$ for the given QP in the BPG codec and achieve a better rate-accuracy trade-off.

\begin{figure}
  \centering
  \includegraphics[width=1\columnwidth]{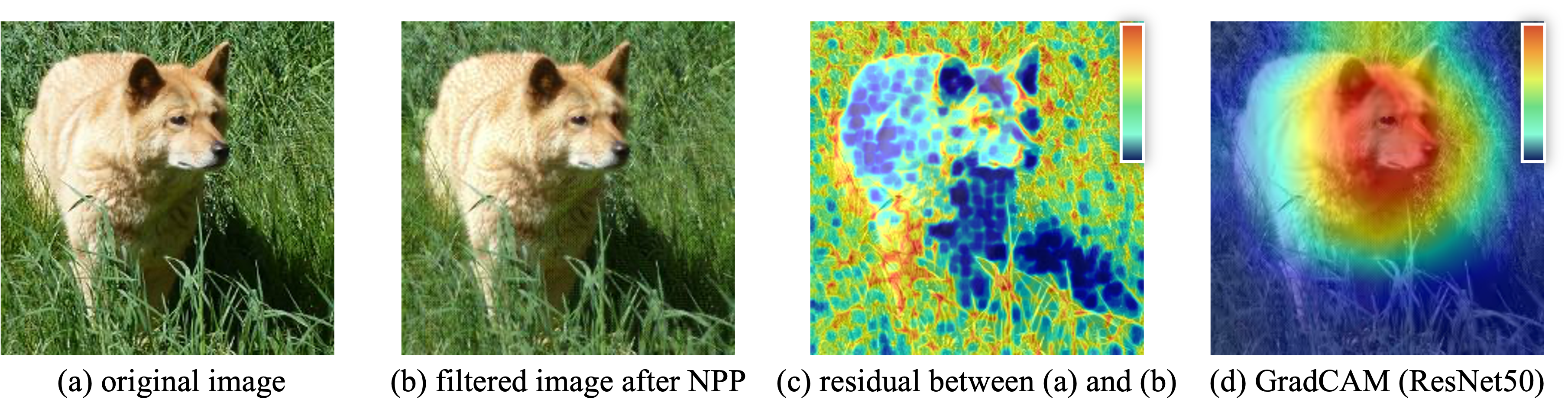}
  \caption{Visualization results of the neural preprocessing module. The color in (c) and (d) represents the values of residual and localized class-discriminative regions form GradCAM~\cite{selvaraju2017gradcam}, where larger values are denoted by red color.}
  \label{fig:cam}
\end{figure}

Here we give an example in Fig.~\ref{fig:cam} to show the effectiveness of our preprocessing module. Fig.~\ref{fig:cam}(a) and (b) represent the original image and output from the NPP module, respectively. Moreover, the corresponding compressed file sizes using the BPG~\cite{BPG}~($QP=37$) codec are 63.7kb and 47.0kb. At the same time, Fig.~\ref{fig:cam}(c) shows that the information discarded by the preprocessing module is mainly distributed in the background region. In contrast, based on the GradCAM method~\cite{selvaraju2017gradcam}, the classification network~\cite{he2016resnet} focuses on the foreground
\textit{Dingo} in the image as shown in Fig.~\ref{fig:cam}(d). 
These results prove that the preprocessing module can preserve more critical semantic information for the downstream analysis tasks and reduce the irrelevant information for bitrate saving.

\subsection{Proxy Network}
\label{sec:proxy}
In our framework, to enable an end-to-end optimization for the whole system,  a learned image compression network is introduced as the proxy network to replace the traditional codec during the backward propagation stage.
Here, we use Minnen's approach~\cite{minnen2018joint} as our proxy network. 

To make sure that the proxy network can well approximate the traditional codec, the reconstruction quality of BPG and Minnen's approach should be similar. Since the learned image compression approach~\cite{minnen2018joint} is optimized based on R-D distortion loss $R + \lambda_p D$ and the quality of the reconstructed image depends on the hyper-parameter $\lambda_p$.
Therefore, we first choose the learned image compression model with a suitable  $\lambda_p$ parameter to approximate BPG codec and then finetune the proxy network~\cite{minnen2018joint} in the following way,
\begin{equation}
    \mathcal{L}_{p}=R_p +\lambda_{p} D= R_p +\lambda_{p} d(\hat{X}, \hat{Y})
\end{equation}\label{eq:proxy_loss}
where $d(\hat{X}, \hat{Y})$ denotes the distortion between the reconstructed image $\hat{X}$ from BPG and the reconstructed image $\hat{Y}$ from the proxy network~(see Fig.~\ref{fig:overview}). $R_p$ represents the corresponding bitrate from the proxy network.
After that, we can obtain an optimized proxy codec to mimic BPG codec.

For the forward stage in the training procedure, we first use the BPG codec to generate the reconstructed image $\hat{X}$, which is input to the analysis models and used to calculate the machine perception loss $\mathcal{D}_{m}$. Then we can obtain the loss function in Eq.~\ref{eq:loss1} based on the bitrate $R_t$ from BPG codec. 
Finally, as shown in Fig.~\ref{fig:overview}, the gradients are calculated and propagated through the paired proxy network to update our preprocessing module in the backward stage.
To end-to-end optimize the preprocessing module based on Eq.~\ref{eq:loss1}, the values of reconstructed image $\hat{Y}$ and the corresponding bitrate $R_p$ from the proxy network are assigned to $\hat{X}$ and $R_t$ from the BPG codec in the back-propagation, respectively.
More implementation details are given in supplementary material.

\begin{figure}[tbp]
	\begin{minipage}[t]{0.498\textwidth}
		\centering
		\includegraphics[width=1.0\textwidth]{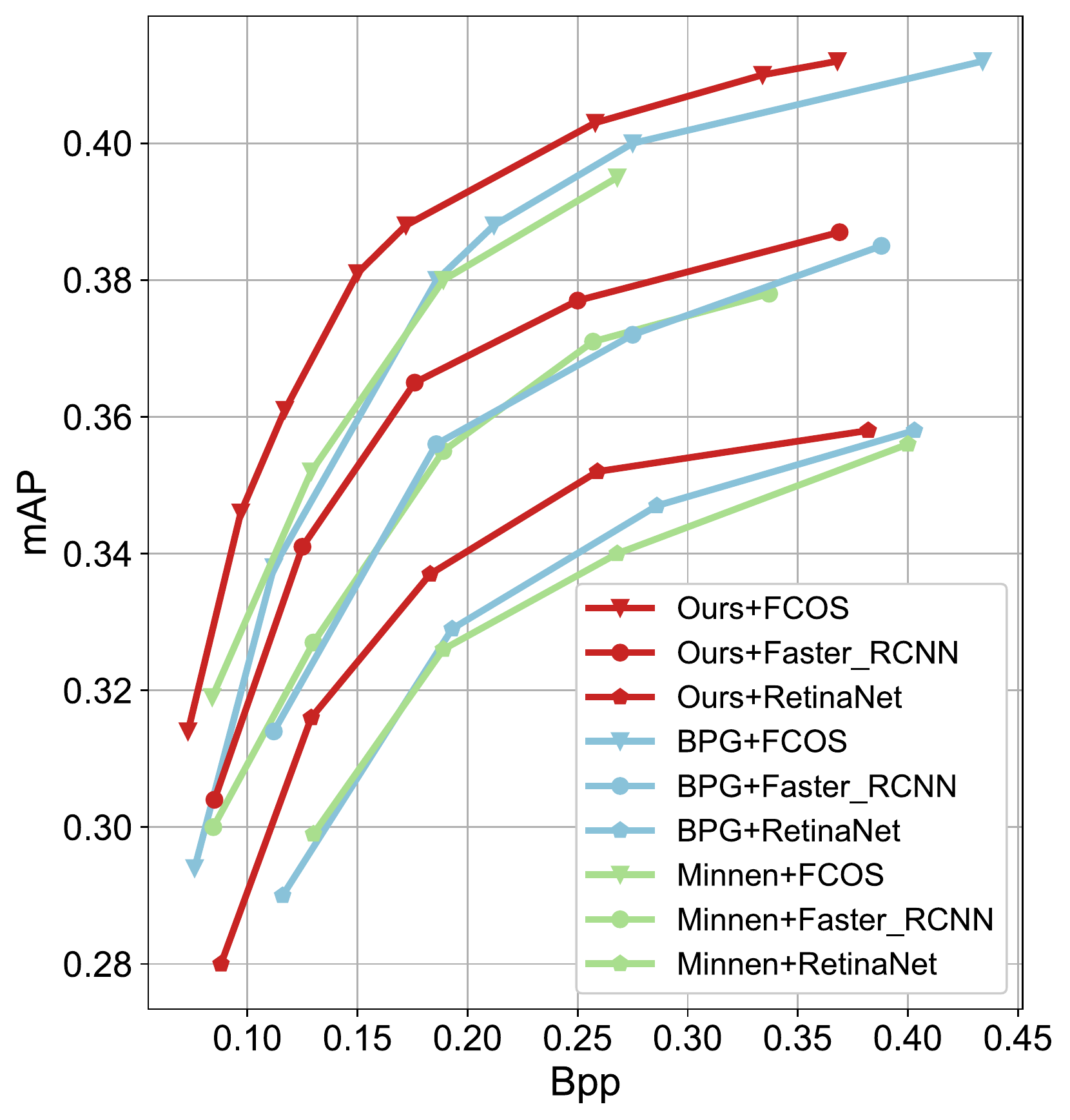}
		\caption{Rate-accuracy(mAP) curves from different compression methods for the object detection tasks on the COCO dataset.}
		\label{fig:detect_results}
	\end{minipage}
	\hspace{0.01\columnwidth}
	\begin{minipage}[t]{0.47\textwidth}
		\centering
		\includegraphics[width=1.0\textwidth]{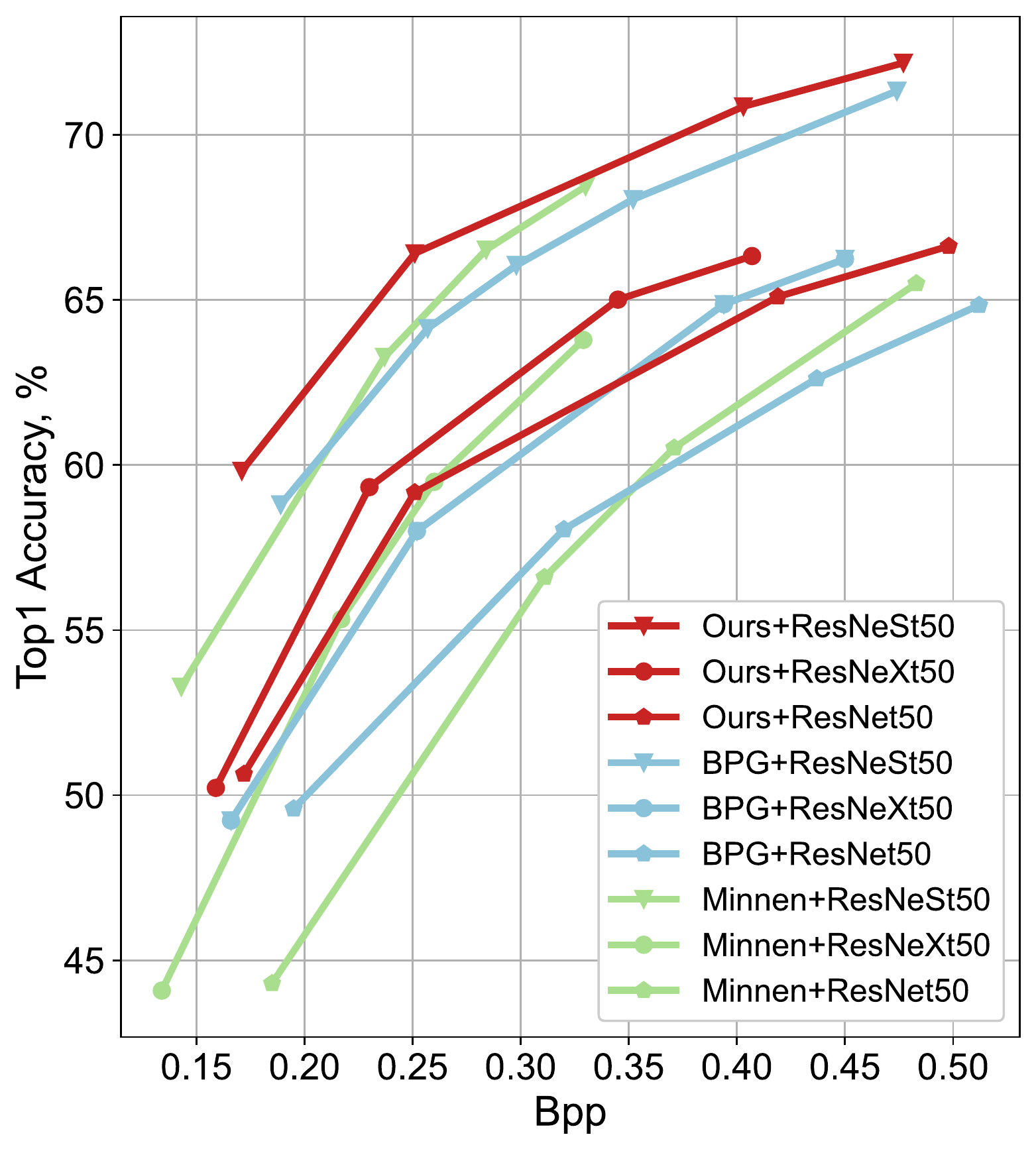}
		\caption{Rate-accuracy(top-1 accuracy) curves from different compression methods for the image classification task on the ImageNet dataset.}
		\label{fig:class_results}
	\end{minipage}
\end{figure}

\section{Experiments}
\subsection{Experimental Setup}

\textbf{Datasets and Backbone Models.}
For the object detection task, we train our framework on the COCO dataset~\cite{lin2014coco}.
We use the COCO \texttt{trainval35k} split (union of 80k images from \texttt{train} and a random 35k subset of images from the 40k image \texttt{val} split) as our training dataset~\cite{cocodataset}.
The mean average precise~(mAP) results are reported by evaluating the proposed framework on the \texttt{minival} split (the remaining 5k images from \texttt{val}) with different compression settings.
In our experiments, three popular object detection baselines FCOS~\cite{tian2019fcos}, Faster-RCNN~\cite{ren2015faster} and RetinaNet~\cite{lin2017RetinaNet} are used for evaluation~\cite{mmdetection}.

For the image classification task, we use the ImageNet dataset~\cite{deng2009imagenet}, which contains 1.28M training images and 50K validation images from 1,000 classes~\cite{imagenetdataset}.
The top-1 accuracy is reported in our experiments. 
To demonstrate the effectiveness of our approach, we use ResNet~\cite{he2016resnet}, ResNeSt~\cite{zhang2020resnest} and ResNeXt~\cite{xie2017resnext} for the performance evaluation~\cite{mmclassification}.

We further evaluate the compression performance in terms of the human visual system by using the perceptual metrics LPIPS~\cite{zhang2018lpips} on the Kodak dataset~\cite{toderici2017full}. We use the bits-per-pixel(bpp) in all experiments to measure the coding cost during the compression procedure.

\textbf{Implementation Details.}
Our whole framework is implemented based on PyTorch~\cite{paszke2019pytorch} with CUDA support and trained on one RTX 3090 GPU card. We use BPG~\cite{BPG} as the traditional codec with different $QP$ values~($QP=28,31,34,37,41$) in our experiments and the corresponding $\lambda$s in Eq.~\ref{eq:loss1} are set as $0.5,1,2,4,8$, empirically. The trade-off parameter $\beta$ is set as 0.5. The weights of the downstream networks such as FCOS~\cite{tian2019fcos} are fixed during the whole training process.

\begin{minipage}[t]{\textwidth}
\begin{minipage}[t]{0.45\textwidth}
\makeatletter\def\@captype{table}
\caption{BDBR~(\%) results between our approach and anchor methods for different backbone networks in object detection task.}
\label{tab:detect}
\resizebox{\textwidth}{11.0mm}{
\begin{tabular}{@{}cccc@{}}
    \toprule
    \multirow{2}*{Anchor} & \multicolumn{3}{c}{Backbones} \\
    \cline{2-4}
    & FCOS & Faster-RCNN & RetinaNet \\
    \midrule
    BPG & -20.3 & -19.5 & -18.8\\
    \midrule
    Minnen & -19.5 & -22.6 & -21.7\\
    \bottomrule
\end{tabular}}
\end{minipage}
\begin{minipage}[t]{0.5\textwidth}
\makeatletter\def\@captype{table}
\caption{BDBR~(\%) results between our approach and anchor methods for different backbone networks in image classification task.}
\label{tab:class}
\resizebox{\textwidth}{11.0mm}{
\begin{tabular}{cccc}
    \toprule
    \multirow{2}*{Anchor} & \multicolumn{3}{c}{Backbones} \\
    \cline{2-4}
    & ResNet50 & ResNeSt50 & ResNeXt50 \\
    \midrule
    BPG & -22.0 & -16.4 & -15.7\\
    \midrule
    Minnen & -24.1 & -12.8 & -14.9\\
    \bottomrule
\end{tabular}}
\end{minipage}
\end{minipage}

The whole training process has the following stages.
First, based on the finetuning procedure in Section~\ref{sec:proxy}, we can obtain several proxy networks that mimic BPG codec with different quantization parameters. 
Then we end-to-end optimize the neural preprocessing module without the quantization adaptive layers according to the loss function in Eq.~\ref{eq:loss1} and set the $QP$ of the BPG codec to a fixed value, such as $QP=34$. 
Finally, we add the quantization adaptive layers into the preprocessing module and further train the preprocessing module by randomly sampling $QP$ values.

Specifically, we use the Adam optimizer~\cite{kingma2014adam} and the initial learning rate is set as 1e-4. The framework is optimized for 400k, 120K and 100k steps during the three training stages.  And the learning rate is reduced to 1e-5 after 320k, 80k and 60k steps when the loss becomes stable. The whole training process takes about five days.

\subsection{Experimental Results}
We compare our preprocessing enhanced image compression method with the existing traditional codec BPG~\cite{BPG} and neural network based compression model~\cite{minnen2018joint}. 
In addition, BD-Rate~\cite{bjontegaard2001calculation}~(BDBR) is used to measure the percentage of saved bitrate with the same accuracy. We use FCOS~\cite{tian2019fcos} and ResNet50~\cite{he2016resnet} as the default backbone networks for object detection and image classification and train the corresponding NPP modules, respectively. 

\textbf{Object Detection}
Fig.~\ref{fig:detect_results} shows the rate-accuracy curve from different backbone networks and compression approaches on the COCO dataset~\cite{lin2014coco}. 
It is obvious that our preprocessing enhanced image compression method shows a much better rate-accuracy trade-off than the baseline approaches on downstream object detection task.
Specifically, compared with the existing BPG codec and learned compression model, our neural preprocessing enhanced codec saves 20.3\% and 19.5\% bitrate at the same mAP value when evaluating on FCOS, respectively.

To further verify the generalization ability of our proposed neural preprocessing module, we perform new experiments by directly applying the NPP optimized for FCOS to other backbone networks like RetinaNet~\cite{lin2017RetinaNet}. 
Experiments in Fig.~\ref{fig:detect_results} show that our compression approach can still outperform the baseline methods and reduces 19.5\% and 18.8\% bitrate when compared with BPG on the downstream  Faster-RCNN~\cite{ren2015faster} and RetinaNet~\cite{lin2017RetinaNet} models, respectively.
These results demonstrate that the proposed solution can be used for downstream networks with different architectures, which is beneficial in practical applications.  
The corresponding BD-Rate results are provided in Table~\ref{tab:detect}. 

\textbf{Image Classification}
We also compare our method with the traditional and learning based codecs on the image classification task.
Fig.~\ref{fig:class_results} shows the rate-accuracy~(top-1) curves from different compression methods on the ImageNet dataset~\cite{deng2009imagenet}. It is noted that our approach still achieves better rate-accuracy performance and saves more than 22.0\% bitrate when compared with traditional codec BPG~\cite{BPG} by evaluating on the ResNet50~\cite{he2016resnet} model.

We further perform new experiments on other mainstream image classification networks by directly using the NPP module optimized for ResNet50. Compared with BPG codec, our framework has 16.4\% and 15.7\% bitrate reduction when the downstream classification networks are ResNeSt~\cite{zhang2020resnest} and ResNeXt~\cite{xie2017resnext}, respectively.
The corresponding BD-Rate results are provided in Table~\ref{tab:class}.

\begin{figure}
    \centering
    \includegraphics[width=0.7\columnwidth]{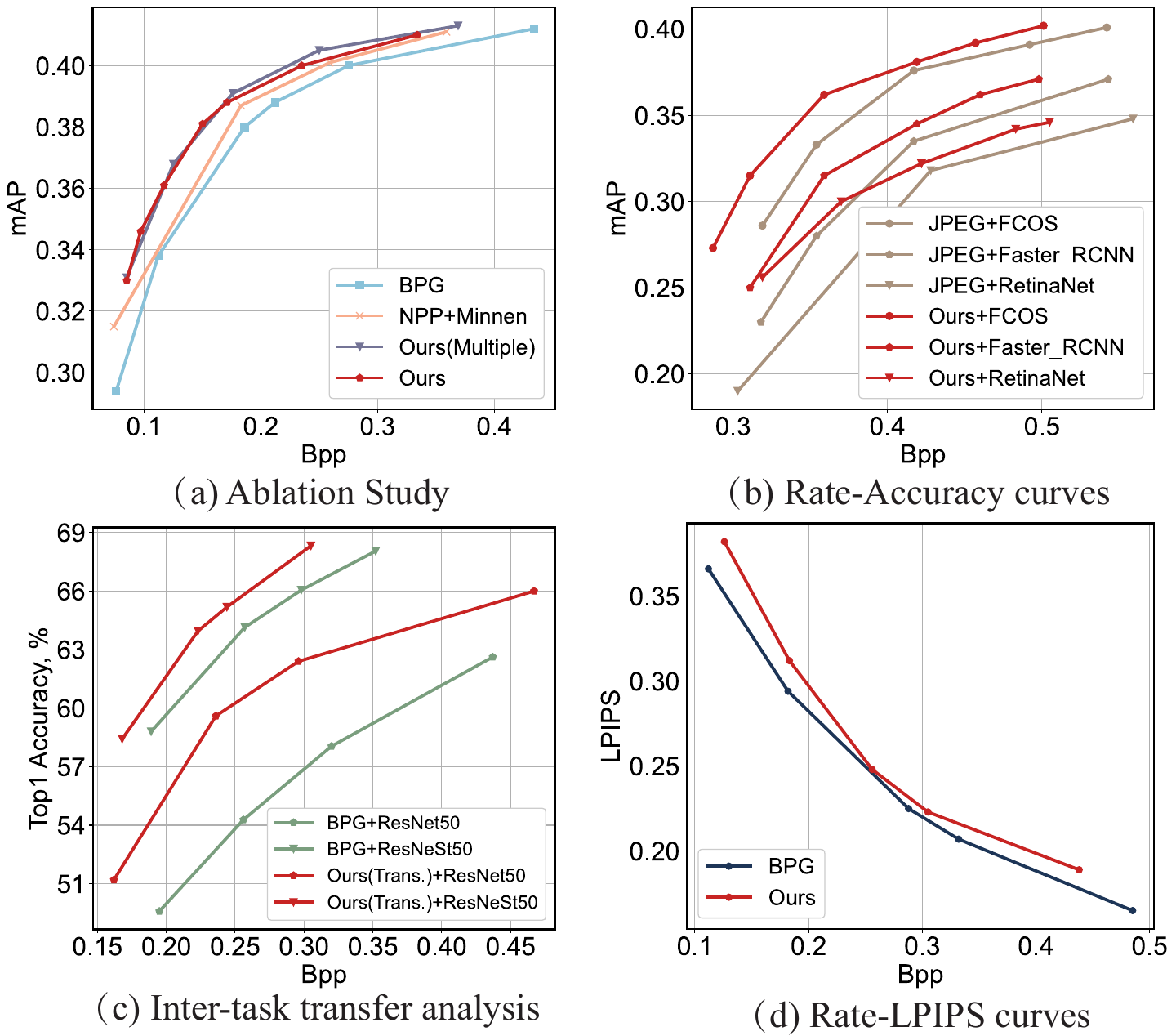}
    \caption{(a) Ablation study. (b) Rate-accuracy curves from our NPP enhanced JPEG codec and native JPEG codecs. (c) Inter-task transfer of our NPP from object detection to image classification. (d) Rate-LPIPS curves from ours approach and BPG codec.}
    \label{fig:ablations}
\end{figure}

\subsection{Ablation Study and Model Analysis}

\textbf{Analysis of End-to-end Optimization.}
In the proposed approach, we use BPG to generate reconstructed images in the forward propagation, while the gradients of the proxy network are used in the backward propagation to update the parameters of the preprocessing module. Here, we also provide the result of directly using the proxy network in both forward and backward propagation, denoted as NPP+Proxy; however, we still use the BPG codec in the inference stage.

Experimental results show that this alternative solution~(NPP+Minnen) can also effectively optimize the preprocessing module and improve the rate-accuracy performance.
As shown in Fig.~\ref{fig:ablations}(a), compared with the original BPG codec, it saves about 14.6\% bitrate at the same mAP value; however, our proposed training strategy is more effective and achieves 20.3\% bitrate saving.
The reason is that our approach uses BPG to generate the reconstructed images in the forward pass, which is consistent with the actual inference stage.

\textbf{Analysis of Quantization Adaptation Strategy.}
Our proposed NPP module is quantization adaptive and can be used for BPG codec with different $QP$s. Here, we provide another alternative solution, \textit{i.e.,} Ours(Multiple), where the quantization adaptive layers are removed and we train different NPP modules for different $QP$s in BPG. Experiments show that it has marginal improvements compared to our quantization adaptive implementation~(See Fig.~\ref{fig:ablations}(a)). However, it needs to train and store multiple NPP models, which brings more storage burden to the encoder side. 

\textbf{Preprocessing Module for JPEG Compression.}
We also provide more experimental results for JPEG compression. Here, we apply the NPP module optimized for the BPG~\cite{BPG} codec to the JPEG~\cite{wallace1992jpeg} without any finetuning.
Experimental results in Fig.~\ref{fig:ablations}(b) show that our proposed preprocessing enhanced JPEG compression achieves more than 8.5\% bitrate savings than the original JPEG codec when evaluating on the FCOS backbone networks. 

\begin{figure}[!t]
  \centering
  \includegraphics[width=1\columnwidth]{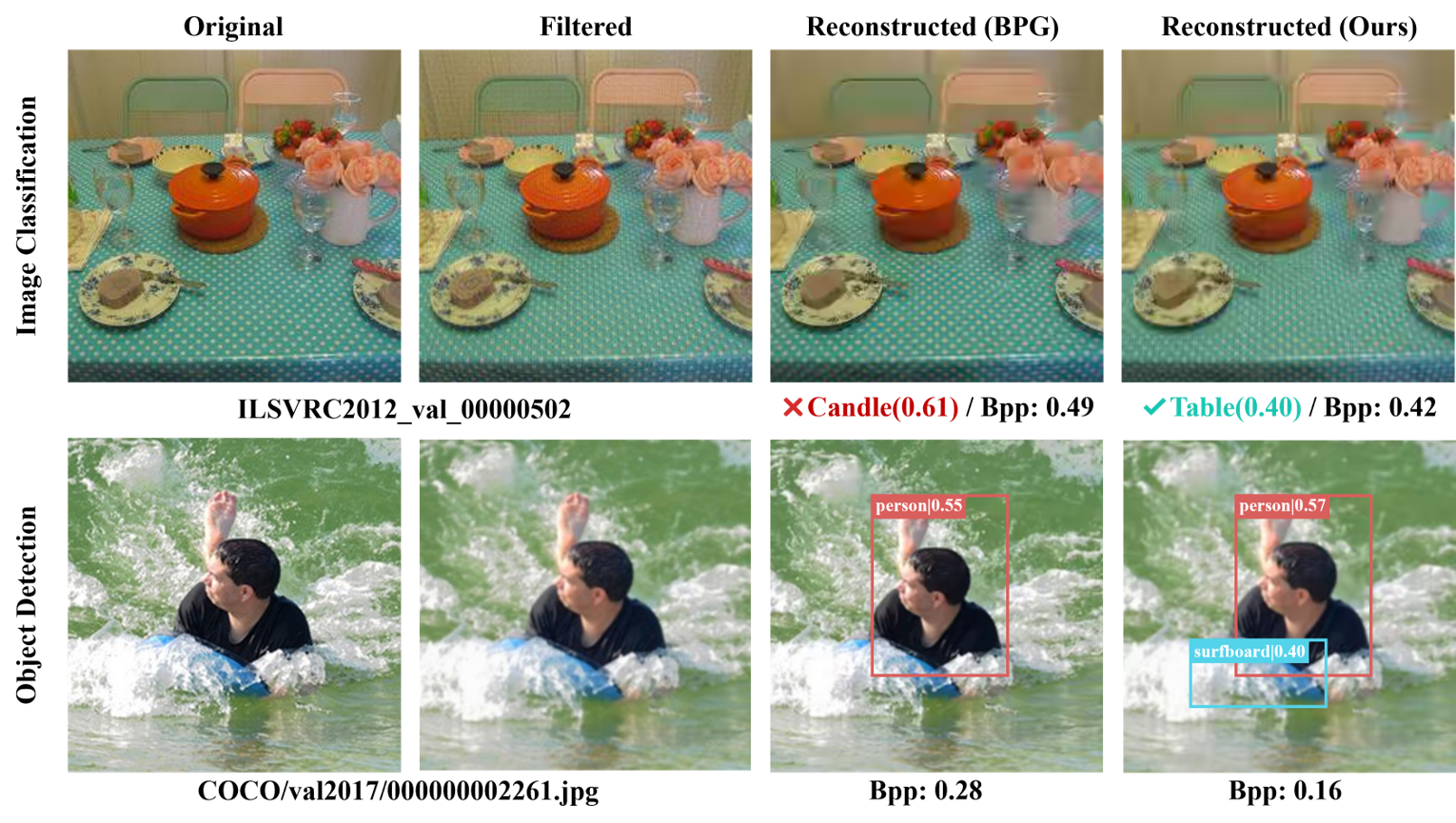}
  \caption{Visualization results of the downstream tasks.}
  \label{fig:qualitative}
\end{figure}

\textbf{Inter-task Transfer of the Preprocessing Model.}
To demonstrate the generalization ability of the preprocessing module, we apply the  preprocessing model optimized for the object detection task to the image classification task.
The results in Fig.~\ref{fig:ablations}(c) show that our transfer method~(Ours(Trans.)) is still useful and achieves more than 10\% bitrate reduction. We have the same observation when we apply the preprocessing model optimized for the image classification task to the object detection task. And more detailed results will be provided in the supplementary material.

\textbf{Compression Performance in terms of Human Visual System.}
We also analyze the compression performance of our preprocessing enhanced image compression approach in terms of the human visual system.
Since our compression framework is optimized for machine vision tasks, the compression performance in terms of PSNR or MS-SSIM drops, which is no surprise. However, when we use more perceptual related metrics like LPIPS~\cite{zhang2018lpips}, we found the gap is narrowing and our approach consumes an additional 8\% bitrate when compared with traditional baseline codec BPG in Fig.~\ref{fig:ablations}(d).  

\textbf{Visualization of Downstream Results.} We provide the visualization results in Fig.~\ref{fig:qualitative} and it is evident that our neural preprocessing module is beneficial for the downstream tasks. For example, the reconstructed images produced by our method in the first row can be correctly classified while the corresponding result from BPG is wrong. At the same time, the proposed also consumes fewer bitrate compared with BPG~(0.42 vs. 0.49).
We have a similar observation for the object detection task in the second row. The small object can be recognized in our compressed results with fewer bitrates while it is missed for the BPG compressed image.

\textbf{Running Time and Complexity.} The number of parameters of our preprocessing module is 9.42M.  For the input image with the size of $224 \times 224$, the inference time of our neural preprocessing module is only 5.17ms, which means it brings little computational complexity to the existing pipeline.

\section{Conclusion}
In this work, based on traditional image compression algorithms, we propose a preprocessing enhanced image compression framework for downstream machine vision tasks. We introduce the neural preprocessing module to achieve a better trade-off between coding bitrate and the performance of machine vision tasks. Furthermore, we propose to use the proxy network to deal with the non-differentiable problem of traditional codec, which 
ensures that the gradients can be back-propagated to the neural preprocessing module. Experiments show that our framework outperforms existing image codecs in object detection and classification tasks. More importantly, our approach shows strong generalization ability for different codecs, backbones, and even for different tasks. 

\textbf{Limitations.}
We only select two representative machine vision tasks(\textit{i.e.,} image classification and object detection) and two codecs(BPG and JPEG) to demonstrate the effectiveness of our approach. Our approach may not apply to other tasks or codecs in practical applications.
However, evaluating more machine vision tasks or codecs will significantly increase the workload, which is beyond the scope of this paper. We will try to include more extensive experiments in the future.   

\textbf{Potential Negative Societal Impacts.}
Our approach is proposed to improve the performance of the downstream tasks. At the same time, the proposed solution can also be used to attack or mislead the downstream tasks by modifying the loss function. 

{
\small
\bibliographystyle{plain}
\bibliography{macros,main}
}

\end{document}